\documentclass[aip,jcp,reprint,showpacs,singlecolumn]{revtex4-1}

\usepackage{amsmath,amsfonts}
\usepackage{graphicx}
\usepackage[acronym]{glossaries}
\usepackage{dcolumn}
\usepackage{bm}
\usepackage{ifpdf}
\usepackage{natbib}
\usepackage{wasysym}
\usepackage[byname]{smartref}
\usepackage{color}
\usepackage{lipsum}
\usepackage{nicefrac}
\usepackage[breaklinks, colorlinks, linkcolor=blue, citecolor=blue, urlcolor=blue]{hyperref}
\usepackage[table]{xcolor}
\usepackage{multirow}
\usepackage{braket}
\usepackage{float}
\usepackage{comment}
\usepackage{enumerate}
\usepackage{enumerate}

\usepackage{amsmath,amssymb,graphicx}
\usepackage{epsfig}

\newcommand{\beq}{\begin{equation}}
\newcommand{\eeq}{\end{equation}}

\newcommand{\bea}{\begin{eqnarray}}
\newcommand{\eea}{\end{eqnarray}}

\begin{document}


\title{Inelastic effects in molecular transport junctions: The probe technique at high bias}

\author{Michael Kilgour}
\affiliation{Chemical Physics Theory Group, Department of Chemistry, University of Toronto,
80 St. George Street Toronto, Ontario, Canada M5S 3H6}

\author{Dvira Segal}
\email{dsegal@chem.utoronto.ca}
\affiliation{Chemical Physics Theory Group, Department of Chemistry, University of Toronto,
80 St. George Street Toronto, Ontario, Canada M5S 3H6}

\date{\today}
\begin{abstract}
We extend the Landauer-B\"uttiker probe formalism for conductances to the high bias regime, and study the
effects of environmentally-induced elastic and inelastic scattering on charge current in single molecule junctions,
focusing on high-bias effects.
The probe technique phenomenologically incorporates incoherent elastic and inelastic effects to the
fully coherent case, mimicking a rich physical environment at trivial cost.
We further identify environmentally-induced mechanisms which generate an asymmetry in the current, manifested as a weak diode behavior.
This rectifying behavior, found in two types of molecular junction models,
is absent in the coherent-elastic limit, and is only active in the case with incoherent-inelastic scattering.
Our work illustrates that in the low bias - linear  response regime, the commonly used
``dephasing probe"  (mimicking only elastic decoherence effects)
operates nearly indistinguishably from  a ``voltage probe" (admitting inelastic-dissipative effects).
However, these probes realize fundamentally distinct  $I$-$V$ characteristics at high biases,
reflecting the central roles of dissipation and inelastic scattering processes
on molecular electronic transport far-from-equilibrium.
\end{abstract}

\maketitle

\section{Introduction}
\label{Sintro}

Electron-nuclei interactions are central to the operation of molecule-based electronic systems \cite{Tao}.
Atomic motion in molecules and their surroundings
opens up new channels for electronic conduction and ultimately,
can result in the emergence of ohmic conduction \cite{diventra,MEbook}.
Energy exchange between conducting electrons and the nuclei
can locally heat the junction \cite{SegalH,DiventraH,Perch,Galp,Lu15}, as was demonstrated
experimentally in Refs. \cite{DiventraE1,DiventraE2,Selzer},
and drive vibrational instabilities \cite{Lu,Dundas,Thoss,SiminePCCP,SimineIF},
even rupturing the junction \cite{lathaRup}. 
These complex many-body effects can be captured within seemingly simple models: The Anderson-Holstein model describes molecular junctions
with a single electronic site coupled to a dominant vibrational mode \cite{GalpRev}. In
Donor-Acceptor junctions, phonon emission and absorption processes facilitate charge transfer between two molecular electronic sites
(or orbitals) \cite{Lu,Dundas,Thoss,Galperin-Park,SiminePCCP,Felix,BijayDA,Thoss2}.
Tight-binding models with several electronic sites and multiple vibrational modes have been employed to describe charge transfer
in extended molecules, linear organic molecules \cite{GalpRev}, and DNA \cite{CunibertiDNA}. 

The Landauer approach offers a simple-exact description of phase-coherent quantum transport \cite{MEbook}.
Given its simplicity, it is appealing to develop it beyond the coherent limit.
Indeed, as was suggested in Ref. \cite{Buttiker}, and expanded on later in Ref. \cite{Pastawski}, one can
simulate incoherent (elastic and inelastic) scattering of electrons with other degrees of freedom, possibly phonons, photons,
and other electrons, by introducing additional terminals, termed probes, into the model system.
The key point in this Landauer-B\"uttiker probe (LBP) technique
is that local charge distributions of the probes
should be set self-consistently such that there is no net average particle current between the
physical system and the probes.
A central advantage of the probe technique is the ability to craft different types of incoherent scattering processes following certain conditions.
These conditions manifest in the ``dephasing" and ``voltage" probes, which respectively allow only elastic or inelastic-dissipative scattering \cite{SalilAB,Kilgour1}.

Previous studies with the LBP technique had demonstrated its utility as a means to
phenomenologically describe low-bias incoherent effects in mesoscopic conductors
\cite{Pastawski,Datta,Dhar,Pastawski14,Meair,Casati1,Casati2,Udo1,Udo2,ButtikerFCS}
and molecular junctions
\cite{Nozaki08,Nozaki12,Pastawski-poly,Chen-Ratner,WaldeckF,Kilgour1,Anantram_linear}
These simulations indicate
that the LBP technique can reproduce several key features in molecular electronic conduction:
A ``Kramers-like'' turnover of conductance as a function of the so-called dephasing strength,
an onset of a thermally activated conductance at high temperature, and a transition from tunneling to ohmic conduction when
increasing the molecular size \cite{Kilgour1}. 
It was also shown in Ref. \cite{Kilgour1} that the LBP method could semi-quantitatively
reproduce experimental conductance results for long, conjugated wires, experimentally examined in Ref. \cite{Frisbie-jacs}.

The implementation of the inelastic LBP (termed voltage probe) to electron and heat
transport problems 
has been typically limited to linear response applications given fundamental and computational challenges.
Only recently, the utility of the voltage probe to high bias problems has been established:
The mathematical uniqueness of the solution was proved in Ref. \cite{Jacquet},
and simulations demonstrating nonlinear charge and energy functionality with a facile convergence were reported in Refs.
\cite{Malay, Tulkki,Sanchez1,Sanchez2,SalilAB, Kilgour2}. While the LBP method is gaining recognition in molecular electronic applications,
the serious fundamental and operational differences between dephasing and voltage probes,
in high bias applications, are generally not recognized.

The rectifier (diode) is one of the simplest nonlinear devices, and a key building block in ordinary semiconductor circuitry.
Rectification has been studied in a variety of junction setups
since the early days of molecular electronics \cite{AviramRatner,MetzgerRev}.
Recent experiments and supporting computational investigations
have demonstrated that one can achieve robust molecular diodes
with rectification ratios of two and even three orders of magnitude in SAMs \cite{Nijhuis15} and single molecules \cite{latha15,Cuevas}.
The mechanism of rectification in many proposed molecular structures is based on the assumption of coherent tunneling, more precisely,
on the fact that deep-tunneling conductance is orders of magnitude smaller than resonant transmission, see for example \cite{vanderzant15,Batista}.
In such tunneling diodes, incoherent thermal effects have been predicted to degrade performance \cite{Kilgour2}.

In this paper, our goal is to simulate environmentally assisted charge transfer in molecular junctions
under large applied voltage biases,
and examine novel effects materializing in this regime due to {\it incoherent inelastic scattering processes}.
Such inelastic effects are assumed to develop from the interaction of electrons with
e.g. intra and inter molecular vibrational degrees of freedom, mimicked here using voltage probes.
Specifically, we address the following questions:
(i) In linear response regime, energy exchange processes are limited, thus
the dephasing and voltage probes act in a similar manner, to realize hopping conduction \cite{Kilgour1}.
What are the unique signatures of inelastic-dissipative scattering processes, vs. elastic processes,
in molecular electronic conduction, at large bias? We will elucidate this point by simulating $I$-$V$ characteristics
of conducting junctions employing the two types of probes, further ``dissecting" the probes and studying their properties.
(ii) In tunneling diodes, environmental-thermal effects degrade operation \cite{Kilgour2}.
Can we identify situations in which the thermal environment enhances, and moreover causes, the nonlinear diode function?
We will suggest two different designs for single-molecule diodes, and demonstrate that while transport is symmetric in the coherent limit,
environmental inelastic effects induce a diode function.

This paper is organized as follows. In Sec. \ref{SMM} we present the LBP method with either elastic effects
or with additional inelastic effects. 
In Sec. \ref{SDV} we examine signatures of these microscopic scattering effects on $I$-$V$ characteristics, far from equilibrium.
In Sec. \ref{Sdiode} we describe two setups in which environmental inelastic effects are essential for realizing the diode effect,
and prove that incoherent elastic processes cannot lead to rectification when the transmission is independent of applied bias (i.e. in the low-bias regime). 
We conclude in Sec. \ref{Ssum}.


\section{Model and Method}
\label{SMM}

We consider a molecule bridging metal electrodes, with vibrational effects introduced via the
probe technique. The total Hamiltonian reads
\bea
\hat H=\hat H_M+\hat H_L+\hat H_R +\hat H_T+ \hat H_P +\hat V_P.
\label{eq:H}
\eea
The molecular electronic states are described
by a tight binding Hamiltonian $\hat H_M$
with $n=1,2,..,N$ sites of energy $\epsilon_n$,
\bea
\hat H_M=\sum_{n=1}^{N}\epsilon_n \hat c_n^{\dagger}\hat c_n
+ \sum_{n=1}^{N-1} v_{n,n+1}\hat c_n^{\dagger}\hat c_{n+1} +h.c.
\label{eq:HM}
\eea
Here, $\hat c_n^{\dagger}$ ($\hat c_n$) are fermionic creation (annihilation) operators of electrons
on each site,  $v_{n,n+1}$ are the inter-site tunneling energies.
Unless otherwise stated we consider identical building blocks, thus we
introduce the short notation $\epsilon_B=\epsilon_n$ and $v=v_{n,n+1}$.
The metal electrodes $\nu=L,R$ are modeled by Fermi seas of noninteracting electrons,
\bea
\hat H_{\nu}=\sum_{k}\epsilon_{\nu,k}\hat a_{\nu,k}^{\dagger}\hat a_{\nu,k},\,\,\,\,\, \nu=L,R.
\label{eq:metal}
\eea
$\hat a_{\nu,k}^{\dagger}$ ($\hat a_{\nu,k}$) are fermionic creation (annihilation)
operators of electrons with momentum $k$ in the $\nu$ lead.
Electrons can tunnel from the $L$ ($R$) metal to site 1 ($N$),
\bea
\hat H_T=\sum_{k} g_{L,k}\hat a_{L,k}^{\dagger}\hat c_1 + \sum_{k} g_{R,k}\hat a_{R,k}^{\dagger}\hat c_N + h.c.
\label{eq:HT}
\eea
%
In the absence of probes, the Hamiltonian dictates phase-coherent electron dynamics.
Incoherent scatterings for electrons on the molecule are introduced
by attaching $N$ internal-fictitious reservoirs to the molecule,
\bea
\hat H_P=\sum_{n=1}^N\sum_{k} \epsilon_{n,k} \hat a_{n,k}^{\dagger}\hat a_{n,k}.
\eea
The $n$th probe exchanges particles with the $n$th site of the molecular wire,
\bea
\hat V_P=\sum_{n=1}^N\sum_{k}g_{n,k}\hat a_{n,k}^{\dagger}\hat c_{n} +h.c.
\label{eq:VP}
\eea
Here $\hat a_{n,k}^{\dagger}$ ($\hat a_{n,k}$) are fermionic creation (annihilation)
operators for an electron in the
$n=1,2,...,N$  probe with momentum $k$, $g_{n,k}$ are the
tunneling energies from the $n$th molecular site into the $n$th probe.
Below we explain the self consistent conditions which the probes should satisfy,
We employ $\nu=L,R$ to identify the (physical) metal electrodes to which the molecule is connected.
The probe terminals are identified by the index $n$, and $\alpha=n,\nu$ counts
all terminals.

The hybridization energy (broadening) of the molecule to the metal leads, $\gamma_{L,R}$,
and its coupling to the internal probes $\gamma_n$ is defined from
\bea
\gamma_{\alpha}(\epsilon)=2\pi\sum_{k}|g_{\alpha,k}|^2\delta(\epsilon-\epsilon_{\alpha,k}).
\label{eq:gamma}
\eea
We work in the wide-band limit, thus take $\gamma_{\alpha}$ as an energy independent parameter.
Unless otherwise stated  we assume that
all sites in the molecule are similarly affected by the environment, and therefore
use a single parameter to denote the probe-molecule hybridization energy,
$\gamma_d\equiv\gamma_n$.
This parameter determines the interaction strength of the electrons with the vibrational environment.
If should be emphasized  that the probe technique not only introduces level broadening (\ref{eq:gamma}),
but further opens up an incoherent transport pathway for electrons  \cite{pathway}.

The model (\ref{eq:H}) does not include interactions thus
the electric current leaving the $L$ contact satisfies the Landauer-B\"uttiker formula,
\bea
I_L=\frac{e}{2\pi\hbar}\sum_{\alpha} \int_{-\infty}^{\infty}\mathcal T_{L,\alpha}(\epsilon) \left[f_L(\epsilon)-f_{\alpha}(\epsilon)\right]d\epsilon.
\label{eq:currL}
\eea
%
$f_{\nu}(\epsilon)=[e^{\beta(\epsilon-\mu_{\mu})}+1]^{-1}$ are the Fermi functions in the physical electrodes, given in terms of the
inverse temperature $k_BT=\beta^{-1}$ and chemical potentials $\mu_{\nu}$.
The distribution functions $f_{n}(\epsilon)$ are determined from the dephasing/voltage probe conditions, explained below.
In direct analogy to Eq. (\ref{eq:currL}), the net current between the $n$th probe and the
physical system can be written as
\bea
I_{n}=\frac{e}{2\pi\hbar}\sum_{\alpha} \int_{-\infty}^{\infty} \mathcal T_{n,\alpha}(\epsilon) \left[f_n(\epsilon)-f_{\alpha}(\epsilon)\right]d\epsilon.
\label{eq:currP}
\eea
The transmission functions in Eqs. (\ref{eq:currL}) and (\ref{eq:currP}) are obtained from
the ($N\times N$) Green's function and the hybridization matrices \cite{diventra},
\bea
\mathcal T_{\alpha,\alpha'}(\epsilon)={\rm Tr}[\hat {\Gamma}_{\alpha'}(\epsilon)\hat G^{r}(\epsilon)\hat \Gamma_{\alpha}(\epsilon)\hat G^a(\epsilon)],
\label{eq:trans}
\eea
where the trace is performed over the $N$ electronic states of the molecule.
The retarded Green's function is
defined in terms of the retarded and advanced Green’s functions,
$\hat G^r(\epsilon)=[\epsilon \hat I-\hat H_M+\frac{i}{2}(\hat \Gamma_L+\hat \Gamma_R +\sum_n\hat \Gamma_n)]^{-1}$,
$\hat G^a(\epsilon)=[\hat G^r(\epsilon)]^{\dagger}$.
The hybridization matrices have a single nonzero value,
\bea
&&[ {\hat \Gamma}_{n}(\epsilon)]_{n,n}= \gamma_{n}(\epsilon),
\nonumber\\
&&[{\hat \Gamma}_{L}(\epsilon)]_{1,1}= \gamma_L(\epsilon), \,\,\,\,
[{\hat \Gamma}_{R}(\epsilon)]_{N,N}= \gamma_R(\epsilon),
\eea
with the molecule-metal hybridization and dephasing strength $\gamma_d$, see Eq. (\ref{eq:gamma}).

The probe technique can be implemented under different self-consistent conditions,
allowing us to craft incoherent electron scattering processes: elastic effects
are implemented via the ``dephasing probe", while dissipative inelastic effects are introduced
through the ``voltage probe".
These probes can be operated in the linear response regime, as well as far from equilibrium \cite{SalilAB}.

\subsection{Dephasing probe}
Incoherent - but elastic - scattering processes are implemented via the dephasing probe condition, where
we demand charge and resolved energy conservation: Each probe is required not only to
conserve the particle number in the junction, but furthermore, it cannot contribute nor dissipate energy to electrons,
at any energy.
Mathematically, this is expressed by requiring the resolved current, the integrand in Eq. (\ref{eq:currP}), to nullify,
\bea 	
I_n(\epsilon)=0, \,\,\,\, n=1,2,...,N.
\label{Dprobe}
\eea
This condition translates into a set of $N$ linear equations for the distributions of each probe
$f_n(\epsilon)$, solved by a matrix inversion at every energy $\epsilon$ within the band (which is taken as broad).

At this point we should clarify on the action of the dephasing probe, since the term ``dephasing" carries
different meanings for different communities.
In the language of molecular orbitals $|M\rangle$, eigenstates of $\hat H_M$, dephasing probes scatter electrons between
orbitals: The dephasing probe absorbs an electron from orbital $|M'\rangle$ and injects it {\it incoherently} elsewhere,
while conserving energy. This type of elastic process is certainly different from
elastic scattering from impurities or the boundaries, a coherent process, accounted for in the original Landauer formula.
The dephasing probe thus opens up a new incoherent-elastic channel for conduction, beyond coherent motion \cite{pathway}.

\subsection{Voltage probe}
Incoherent inelastic-dissipative effects can be introduced
by requiring the net {\it total} particle current flowing
between each probe and the system, Eq. (\ref{eq:currP}), to vanish,
\bea
I_n=0, \,\,\,\,\ n=1,2,..,N.
\label{eq:Vprobe}
\eea
In this case, we force the probe distribution functions to take the form of Fermi functions,
$f_n(\epsilon)=[e^{\beta(\epsilon-\mu_n)}+1]^{-1}$,
and we search for the set $\mu_n$ which fulfills Eq. (\ref{eq:Vprobe}).
In the linear response regime this condition translates to a set of $N$ linear equations which
can be solved formally-analytically to give the chemical potential $\mu_n$ of the probes.
In this limit, energy exchange with the environment (probes) is small, and effectively,
the voltage probe acts similarly to a dephasing probe \cite{Kilgour1}.
Far from equilibrium, voltage probes {\it inelastically} scatter electrons
between molecular orbitals, absorbing electrons, then injecting them with
a range of energies determined by the distribution function within each probe.

Outside the limits of linear response
we need to solve Eq. (\ref{eq:Vprobe}) without any approximations. This equation constitutes a set of $N$ nonlinear equations,
thus an exact analytic solution is in general not forthcoming. As discussed in Ref. \cite{Kilgour2},
we address this problem by retracting to a fully numerical procedure.
We employ the Newton-Raphson method to find the unique set of roots for
the $N$ probes' chemical potentials $\mu_n$ \cite{Jacquet,SalilAB}.
For example, the 3-probe case iterates according to
\bea
\begin{bmatrix}
\mu_1^{k+1} \\ \mu_2^{k+1}\\ \mu_3^{k+1}
\end{bmatrix}	
=
\begin{bmatrix}
\mu_1^k\\ \mu_2^k\\ \mu_3^k
\end{bmatrix}
-
\begin{bmatrix}
\ \frac{\partial I_{1}}{\partial \mu_1} && \frac{\partial I_{1}}{\partial \mu_2} && \frac{\partial I_{1}}{\partial \mu_3} \\ \frac{\partial I_{2}}{\partial \mu_1} && \frac{\partial I_{2}}{\partial \mu_2} && \frac{\partial I_{2}}{\partial \mu_3}\\ \frac{\partial I_{3}}{\partial \mu_1} && \frac{\partial I_{3}}{\partial \mu_2} && \frac{\partial I_{3}}{\partial \mu_3}
\end{bmatrix}^{-1}_k
\begin{bmatrix}
\ I_{1} \\ I_{2}\\ I_{3}
\end{bmatrix}_k
\nonumber\\
\eea
Here, $\mu_n^k$ denotes the potential profile in the $n$th probe after the $k$th iteration, beginning from a certain-physical initial guess.
The method converges regularly for a wide range of parameters, with certain conditions
requiring specialized guesses.
Convergence is confirmed by checking that $\mu_n^k$ remains fixed with iterations,
and that the leakage current to each probe
is small ($|I_n|/|I_{L}|<10^{-5}$).
Using either (dephasing, voltage) probe condition the current leaving the $L$ contact should be identical to the current
reaching the $R$ terminal, $I_L=-I_R$. However, the energy resolution is different
as we demonstrate below.

\begin{figure*}
\vspace{0mm}
{\hbox{\epsfxsize=180mm \epsffile{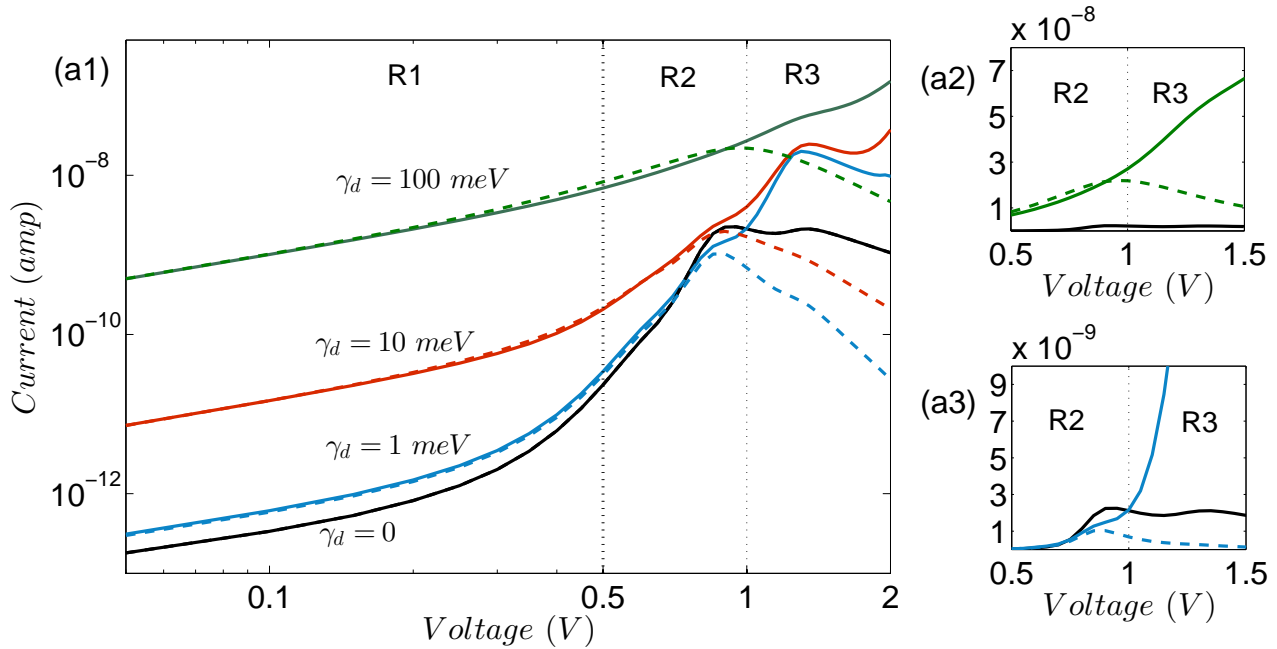}}} 
\caption{
(a)
$I$-$V$ characteristics with a linearly deformed bridge
[Eq. (\ref{eq:profile})] with $N$=6 sites of height $\epsilon_B=0.5$.
Other parameters are $v=0.05$, $\gamma_{L,R}=0.2$, $\gamma_d$=0, 0.001, 0.01, 0.1 eV
(organized bottom to top at low voltage),
with dephasing probe (dashed) and voltage probe (full), $T=298$ K.
The dotted line at $\Delta\mu=0.5$ eV marks the value beyond which molecular resonances appear within the bias window.
Beyond $\Delta\mu=1.0$ eV, more than half of the sites of the bridge are situated within
the bias window. Zoom over (a2) $\gamma_d=100$ meV, (a3) $\gamma_d=1$ meV.
}
\label{FigIV}
\end{figure*}

\begin{figure*}
\vspace{0mm} \hspace{0mm}
{\hbox{\epsfxsize=180mm \epsffile{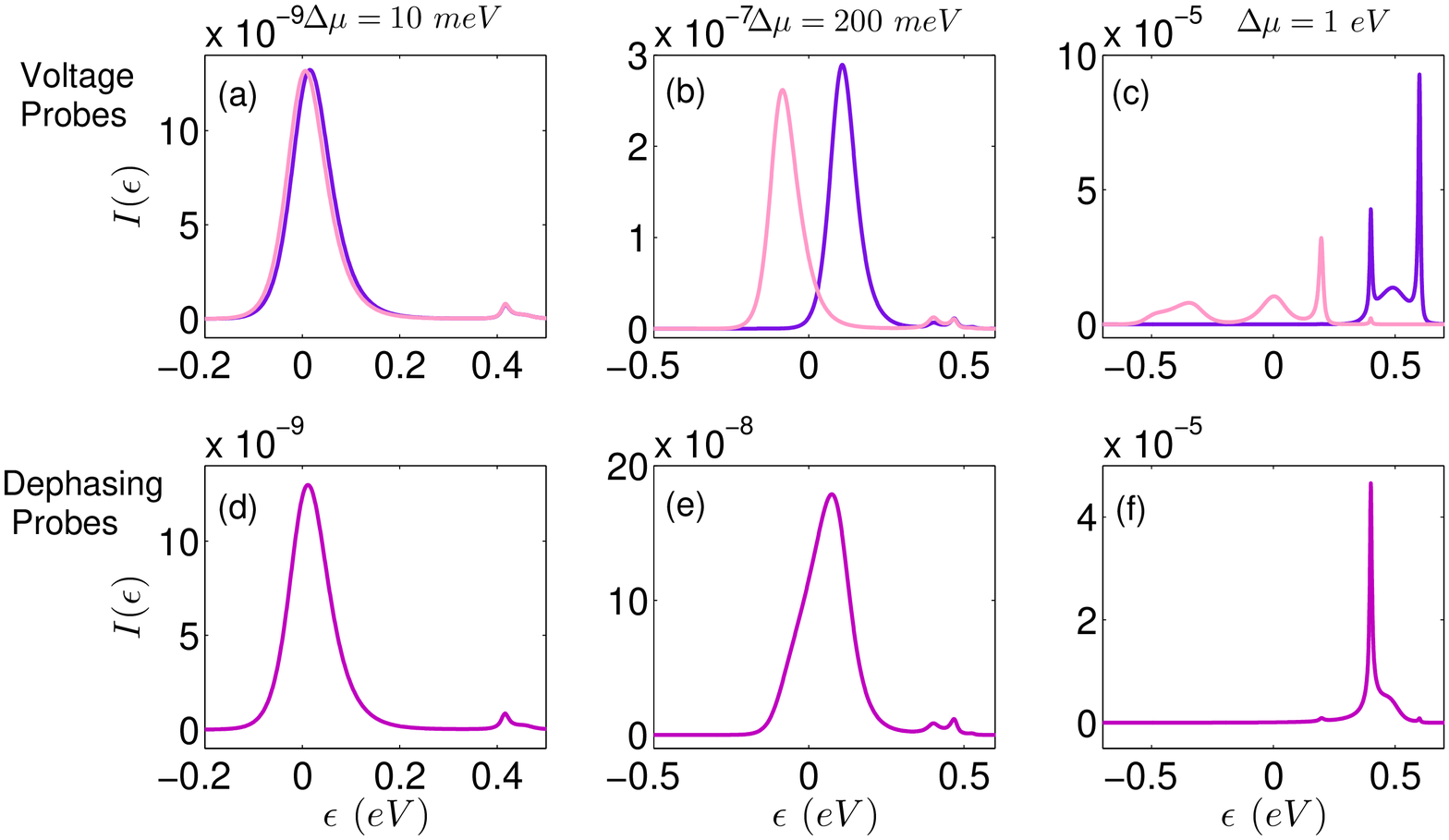}}}  
\caption{Magnitude of energy-resolved currents at the left and right contacts, calculated
as the integrand of Eq. (\ref{eq:currL}), and an analogous current at the right side.
The voltage probe (top) permits energy dissipation,
therefore the energy-resolved current leaving the left-high bias-lead (dark purple) is not the same as that arriving at the right low-bias lead (pink).
The dephasing probe (bottom) does not allow dissipation, and the left and right energy-resolved currents are identical.
Parameters are the same as in Fig. \ref{FigIV}, $\gamma_d=10$ meV.
(a) and (d): Low voltage with off-resonant conduction.
(b) and (e): Low-intermediate voltage with off-resonant conduction, and
(c) and (f): High-bias simulations with resonant transmission.}
\label{FigP2}
\end{figure*}

\section{B\"uttiker's probes at finite bias}
\label{SDV}

\subsection{Current-voltage characteristics}


The linear response conductance of the model (\ref{eq:H})-(\ref{eq:VP}) was recently analyzed in Ref. \cite{Kilgour1}.
In the absence of probes, coherent electrons either ballistically conduct (in resonance situations),
or via the tunneling mechanism (when $\epsilon_B>\gamma_{\nu}$, $v$, $T$ ).
It was shown that the significant contribution of the probes has been to support a hopping conductance beyond tunneling, satisfying
\bea
G_{H}&\sim& G_0
A(T) \frac{v^2}{\epsilon_B^4} \frac{\gamma_d^2}{ N + l }.
\label{eq:TH}
\eea
Here $G_0=e^2/2\pi\hbar$ is the quantum of electrical conductance per spin,
$A(T)$ is a dimensionless temperature-dependent prefactor (typically weaker than an Arrhenius factor),
$N$ is the number of sites on the chain, and $l$ a parameter characterizing the crossover from tunneling to hopping.

In the low bias limit, dephasing and voltage probes similarly affect electronic conductance \cite{Kilgour1}.
In contrast, Fig. \ref{FigIV} shows that far from equilibrium the two probes arrange for very different behavior due to energy exchange processes
which are allowed under the voltage probe condition.
We perform our simulations assuming a linear-ramp potential profile,
\bea
\epsilon_n= \epsilon_B+ \frac{\Delta \mu}{2} - \frac{\Delta \mu (n-1)}{N-1},
\label{eq:profile}
\eea
which corresponds to a barrier being symmetrically reshaped from
a rectangular into a triangular form by the applied voltage.
The $1$st and $N$th sites are pinned at a height $\epsilon_B$ from the electrodes'
Fermi energies $\mu_L$ and $\mu_R$, respectively.
The remainder of the wire sites follows a linearly decreasing profile.

Fig. \ref{FigIV} displays several important properties in different regimes
of the bias voltage, identifies by regions R1, R2 and R3:

(R1) In the low-bias region R1, $\Delta\mu\ll\epsilon_B,k_BT$, conduction is off-resonant in the sense that all
molecular orbitals (and thus transmission resonances)  are placed outside the bias window.
In this region, the two probes (full and dashed lines) similarly act on the junction.
The effect of the environment is to promote an incoherent hopping conduction,
monotonically enhancing the current with $\gamma_d$ in this range of parameters \cite{Kilgour1}.
The current as well increases with voltage since molecular levels are being gradually tuned down,
to shrink the energy gap $\epsilon_B$ between the bridge and the chemical potentials of the metals.

(R2)
At high bias, once molecular electronic states move into the bias window,
the two probes begin to affect transport distinctly.
Within our parameters, the bridge is tilted into a resonance condition
once $\epsilon_6\leq\mu_L$, satisfied here when $\Delta \mu \geq 0.5$ eV. 
More molecular states are pushed into the bias window as we continue to increase the bias.
Eventually, when $\Delta\mu=1$ eV,
three molecular electronic sites (out of six) already satisfy the resonant condition.
We identify the region beyond that, $\Delta \mu>1$ eV, as the ``deep resonant regime" R3,  to be discussed below.

As we highlight in panels (a2) and (a3), in the R2 region the probes act similarly around $\Delta \mu\sim 0.5$ eV.
As we approach region R3, significant deviations start to show up particularly at
low dephasing strengths $\gamma_d=1,10$ meV. It is interesting to note that the probes more similarly act at higher dephasing strength. This appears to be the case due to the fact that at high dephasing strength the incoherent current is very large relative to the direct L-R tunneling, and extreme level broadening due to $\gamma_d$ dictates performance.

(R3) In the high bias region $\Delta\mu\gtrsim 1.0$ eV, the dephasing probe {\it suppresses} the resonant current,
in some cases even {\it below} the coherent limit, while the voltage probe generally supports increasingly high currents (negative differential conductance may show up, but the overall trend is the growth of current with voltage).
This result is one of the main observations of this paper: Under large bias, elastic decoherence effects
degrade the current in uniform bridges, but energy exchange processes facilitate conduction.
The explanation for this behavior is straightforward:
The bridge energetics shift significantly under large biases assuming the profile (\ref{eq:profile}),
or a similar deformation.
When energy exchange processes are allowed, electrons cross the junction while dissipating energy within each site,
resonantly following the molecular electronic energy profile, to yield high currents.
In contrast, since the dephasing probe condition does not allow energy exchange, at high bias
electrons need to cross a junction with highly dispersed energy states---elastically. This scenario leads to low currents.
Thus, the same potential profile can be beneficial for conduction if electrons can ``slide down", while leaving excess energy behind, or
disadvantageous, when only elastic processes are permitted.

For the dephasing probe, the precise value of voltage where the turnover behavior takes place depends
on the model assumed for the potential profile.
However, the overall crossover behavior is robust to the details of the potential drop,
as long as levels are monotonically shifting away from equilibrium in response to bias.

We also comment that even-odd effects show up in the $I$-$V$ characteristics when $\gamma_d=0$ (not shown here):
When $N$ is odd, the molecular site at the center of
 the bridge, $n=(N+1)/2$, precisely overlaps with the
Fermi energy of the left lead at high bias, resulting in a large coherent contribution relative to the even-$N$ setup.


We emphasize that the decay of the current with bias with dephasing probes,
as observed in Fig. \ref{FigIV}, stems from the
dispersion of molecular energies in the high-bias regime.
At equilibrium, the molecular-backbone (bridge) analyzed here includes
identical sites; under bias, Eq. (\ref{eq:profile}) dictates level separation.
In contrast, if we were to begin with an energetically asymmetric junction, e.g., a two-site molecule with $\epsilon_1<\epsilon_2$,
levels would shift  toward each other when $\mu_L>\mu_R$ [again, assuming (\ref{eq:profile})],
and the two probes (voltage, dephasing) would then similarly affect the current, enhancing it through the hopping contribution
\cite{Kilgour2}.


Can experiments reflect the significant deviation between probes in region R3 as reported in Fig. \ref{FigIV}?
Inspecting the literature we note that in relevant setups measurements are typically reported up to $\sim$1 V,
covering regions R1 and R2. Specifically,
$I$-$V$ characteristics of conjugated multi-unit linear wires generally exhibit an
enhancement of current with voltage, more consistent with the voltage probe results.
For example, the OPI and ONI self-assembled monolayers of Refs. \cite{Exp-Frisbie, Frisbie-jacs}
and OPE wires in Ref. \cite{Exp-Lu}
were measured with bias up to 1.5 V demonstrating a current monotonically increasing with bias, with an estimated barrier height $0.3-0.7$ eV. These values are in line with parameters assumed in our simulations.
Similar results were reported in ultra-thin CuPc (Copper Phthalocyanine) heterojunctions \cite{Bufon}.
Energy dissipation is expected to take place in these systems, consistent with the behavior of the voltage probe.
According to our simulations, beyond 1.5 V (not reported in these experiments), the majority of electronic levels are placed within the bias window and inelastic effects become highly influential.
In this case, unless energy dissipation from the conducting system to the environment is highly effective,
significant heating will take place, leading to junction's instabilities and eventually to rupture.

The transition from  ``direct tunneling" to ``field emission
transport" can be revealed from a Fowler-Nordheim (FN) plot, looking at $\log(I/V^2)$ as a function of $1/V$,
to extract the barrier height from the minima of this plot.
Within our parameters, this analysis is relevant in the region $\Delta \mu < 1$ eV---where the two probes similarly act.
Thus, a typical FN plot (as reported e.g. in  Refs. \cite{Exp-Frisbie, Frisbie-jacs}) does not visibly reveal signatures of inelastic effects.

\subsection{Inside the probes}

When voltage probes are attached to the molecular bridge, dissipation takes place stepwise at each site;
the probes absorb electrons at a certain energy, and re-inject them, typically at lower energies. Dephasing probes, in contrast,
conserve energy. This fundamental distinction is illustrated in Fig. \ref{FigP2}
where we compare (absolute value) energy-resolved currents at the left and right contacts,
the integrand of Eq. (\ref{eq:currL}), and the analogous one at the right end.

Panels (a) and (d) display the behavior at low bias, $\Delta\mu=$ 10 meV, using voltage and dephasing probes, respectively.
The voltage probe condition shows little separation between $|I_L(\epsilon)|$ and $|I_R(\epsilon)|$, thus close to zero dissipation.
The dephasing and voltage probes here act nearly identically, and produce almost equivalent resolved currents.
As we increase the bias to $200$ meV in panels (b) and (e), we start to see peak separation in resolved currents under the voltage probe,
and the voltage and dephasing probe resolved currents are now distinguishable.
However, in this regime both probes still produce very similar total currents, since
critically, individual molecular resonances have not yet entered the bias window.
At high bias, $\sim 1$ eV, voltage probe dissipation has become large,
with $|I_L(\epsilon)|$ and $|I_R(\epsilon)|$ not overlapping at all in panel (c).
We can also identify several distinct peaks corresponding to molecular resonances.
The dephasing probe  in panel (f) shows a much sparser picture, with a single, very weak resonance in the resolved current.
This produces the result shown in Fig. \ref{FigIV}, where the voltage probe current continues
to increase in this regime, diverging from the collapsing dephasing probe current.


\begin{figure}[tbp]
\vspace{0mm} \hspace{0mm}
{\hbox{\epsfxsize=90mm \epsffile{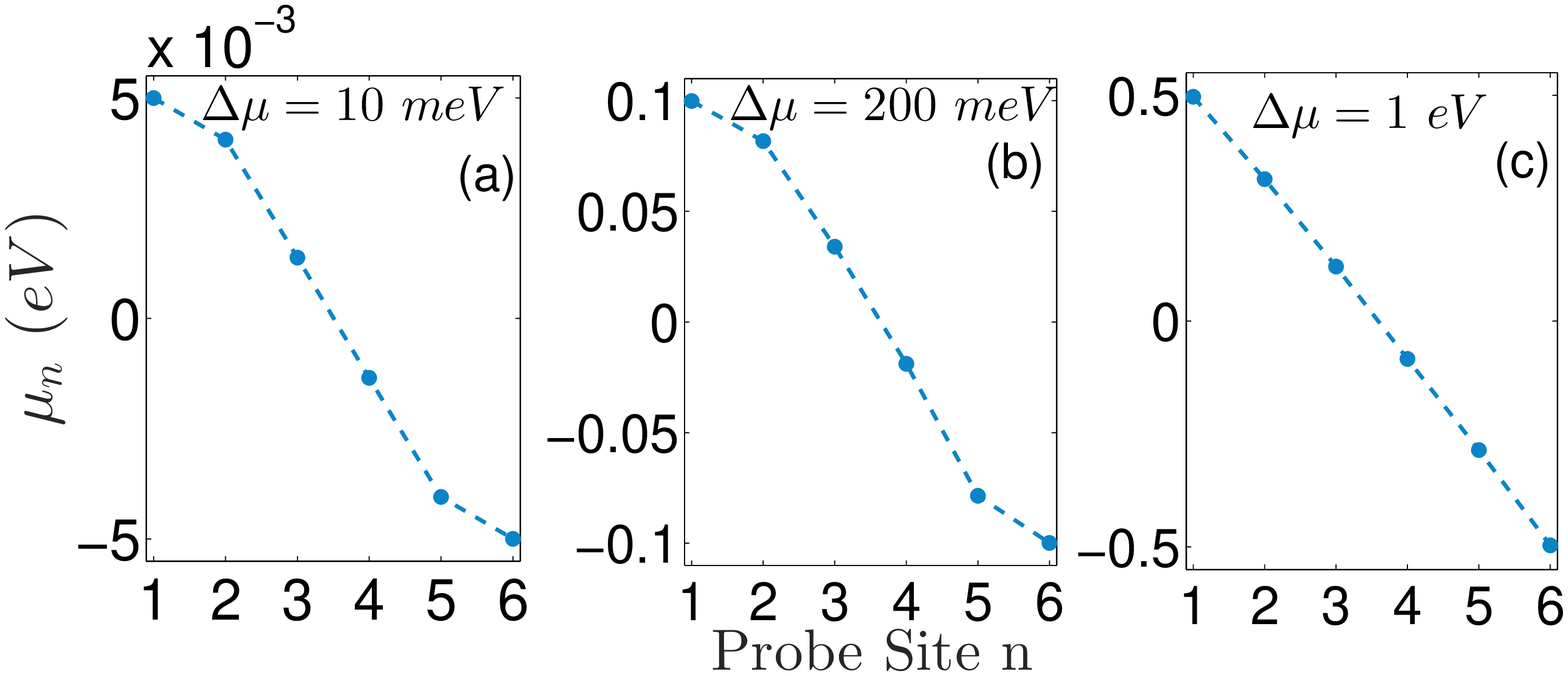}}}  
\caption{Potential profile within voltage probes along the chain
for $\gamma_d=10$ $meV$ in the R1, R2, R3 regions for (a), (b), (c), respectively.
Parameters are the same as in Fig. \ref{FigIV}.}
\label{FigP1}
\end{figure}

We further analyze the operation of the voltage probe in different bias regimes by
plotting in Fig. \ref{FigP1} the self-consistently calculated chemical potential
$\mu_n$ within each probe along the chain. Recall that the probes' electron distributions
follow the form of a Fermi function with a fixed temperature $T$.
In the case of the linearly deformed bridge as used here, the probes' potential profile is close to being linear at high bias.
The reason is that molecular resonances separate as the bias increases, and the probes follow this dispersion.
In the case of a flat molecular bridge (un-deformed by applied bias), it can be shown that the internal profile at high bias becomes highly nonlinear, with probes' internal potentials clustering near dominant molecular resonances.

\section{Environmentally induced diodes}
\label{Sdiode}

With the machinery in hand to explore inelastic effects at finite bias,
we turn our attention to one of the fundamental building blocks of nanoelectronics,
a single-molecule diode. The commonly examined ``tunneling diode" relies on the dramatic contrast between deep
tunneling conductance and ballistic transport, see for example recent experiments \cite{Nijhuis15,Cuevas} and theoretical studies \cite{vanderzant15,Batista}. In these systems, rectification arises e.g. out of an energetic asymmetry
in the bridge - sensitive to the bias polarity - with decoherence and inelastic effects shown to be
destructive to performance \cite{Kilgour2}.

Here, in contrast, we focus on a different mechanism for the diode effect,
originating solely from the interaction of electrons with the environment.
This environmentally-induced diode (EID) effect is relatively weak in our modeling,
but may provide insight into weak rectification in systems which are expected to be symmetric coherent conductors.
To isolate the effect, we maintain the energies of molecular states fixed under bias.
Obviously, in real systems various effects may play together to yield an asymmetric current.

We present below two types of EIDs. In both cases a spatial asymmetry is included,
but the current is symmetric (no rectification) in the Landauer coherent limit and under dephasing probes.
However, rectification appears when incoherent {\it inelastic} effects are allowed.
In type-1 EIDs, the molecule is asymmetrically hybridized to the leads, see Fig. \ref{FigD1}.
Type-2 EIDs are structurally uniform, but the coupling strength to the environment varies asymmetrically along the molecule,
see Fig. \ref{FigD2}.

\subsection{Rectification via asymmetric metal-molecule coupling}
\label{D1}

\begin{figure}[tbp]
\vspace{0mm} {\hspace{5mm}\hbox{\epsfxsize=60mm \epsffile{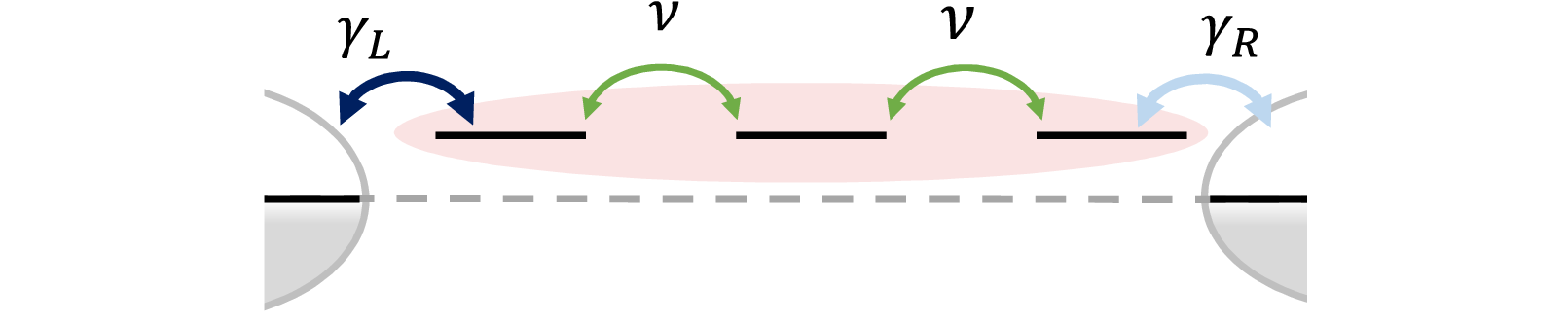}}} 
\hspace{3mm}
{\hbox{\epsfxsize=89mm \epsffile{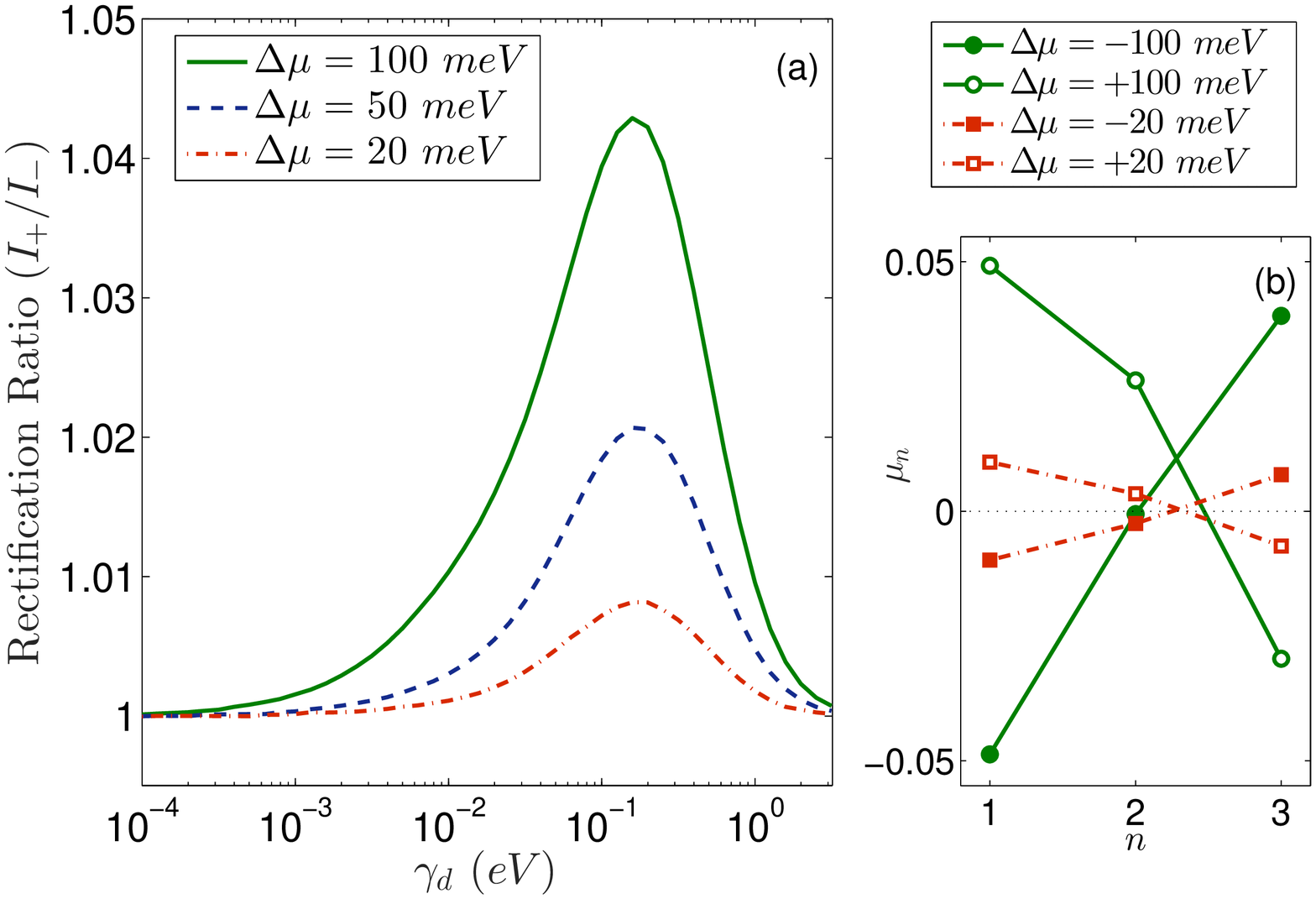}}}  
\caption{
Type-1 EID. Top: Schematic representation of a diode induced by inelastic effects.
The electronic structure is symmetric, but hybridization strengths to the leads are different, illustrated by
arrows of different color at the edges.
(a) Rectification ratio as a function of environmental coupling.
(b) Potential profile within the voltage probes
demonstrating the development of an asymmetry in the junction's response
to applied bias at peak value $\gamma_d=0.16$ eV.
Parameters are $\epsilon_B=0.3$, $v=0.05$, $\gamma_{L}$=0.2 eV, $\gamma_L=10\times \gamma_R$, $T=298$ K and $N$=3, voltage probe condition.
The dotted line identifies the symmetry axis.
}
\label{FigD1}
\end{figure}

We consider a junction with a symmetric molecular backbone but dissimilar couplings to the electrodes, $\gamma_L\neq\gamma_R$.
Assuming the bridge energies do not shift under bias, in the fully coherent Landauer picture
this setup leads to identical values for the forward and reversed currents, thus zero rectification.
We now include identical probes along the wire.
Rectification is missing if we use dephasing probes, see Sec. \ref{SecND}.
However, once we allow inelastic effects (voltage probes), we find
that the magnitude of the current is higher from the strongly coupled contact to the weak one,
than in the reverse direction, see Fig. \ref{FigD1}.

The effect can be correlated with the behavior of the potential profile within probes.
As we point out below in Sec. \ref{SecND}, close to equilibrium the probe potential profile
satisfies $\mu_n^++\mu_n^-\sim\mu_L+\mu_R$, with little rectification showing up.
Note that according to our conventions
$\mu_L+\mu_R=0$, marked by a dotted line in Fig. \ref{FigD1}(b) and
$\mu_n^{\pm}$ is the probes' profile under a forward ($+$) and backward ($-$) polarity.
Inspecting Fig. \ref{FigD1}(b),
at high bias this mirror symmetry is obviously violated,  $\mu_n^++\mu_n^->0$, manifested by the diode effect.

Fig. \ref{FigD1}(a) displays a crossover behavior for rectification:
In short chains and at low $\gamma_d$, dissipationless tunneling dominates transport.
Rectification gradually increases with growing $\gamma_d$ once inelastic conduction becomes substantial.
However, at high enough $\gamma_d$ peak broadening makes forward and backward currents similar,
to dampen rectification.
As expected, the diode effect grows with applied voltage bias, as dissipation effects become more substantial.
However, to be consistent with the assumption of fixed bridge energetics irrespective of $\Delta \mu$,
we limit our simulations here to relatively low biases.



\subsection{Rectification with spatially asymmetric molecule-bath interactions}
\label{D2}

\begin{figure}[tbp]
\vspace{0mm}
{\hspace{5mm}\hbox{\epsfxsize=60mm \epsffile{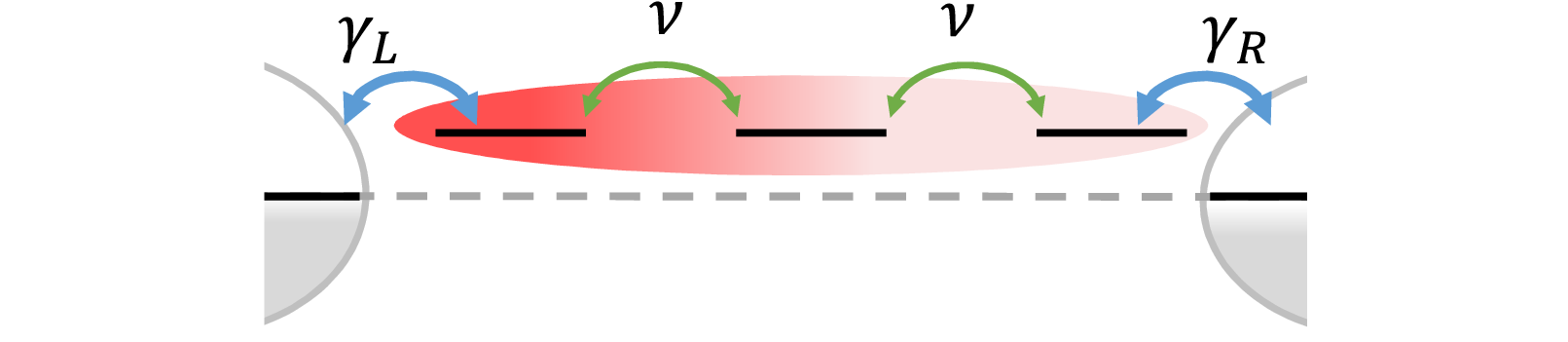}}}
\hspace{-4mm}{\hbox{\epsfxsize=90mm \epsffile{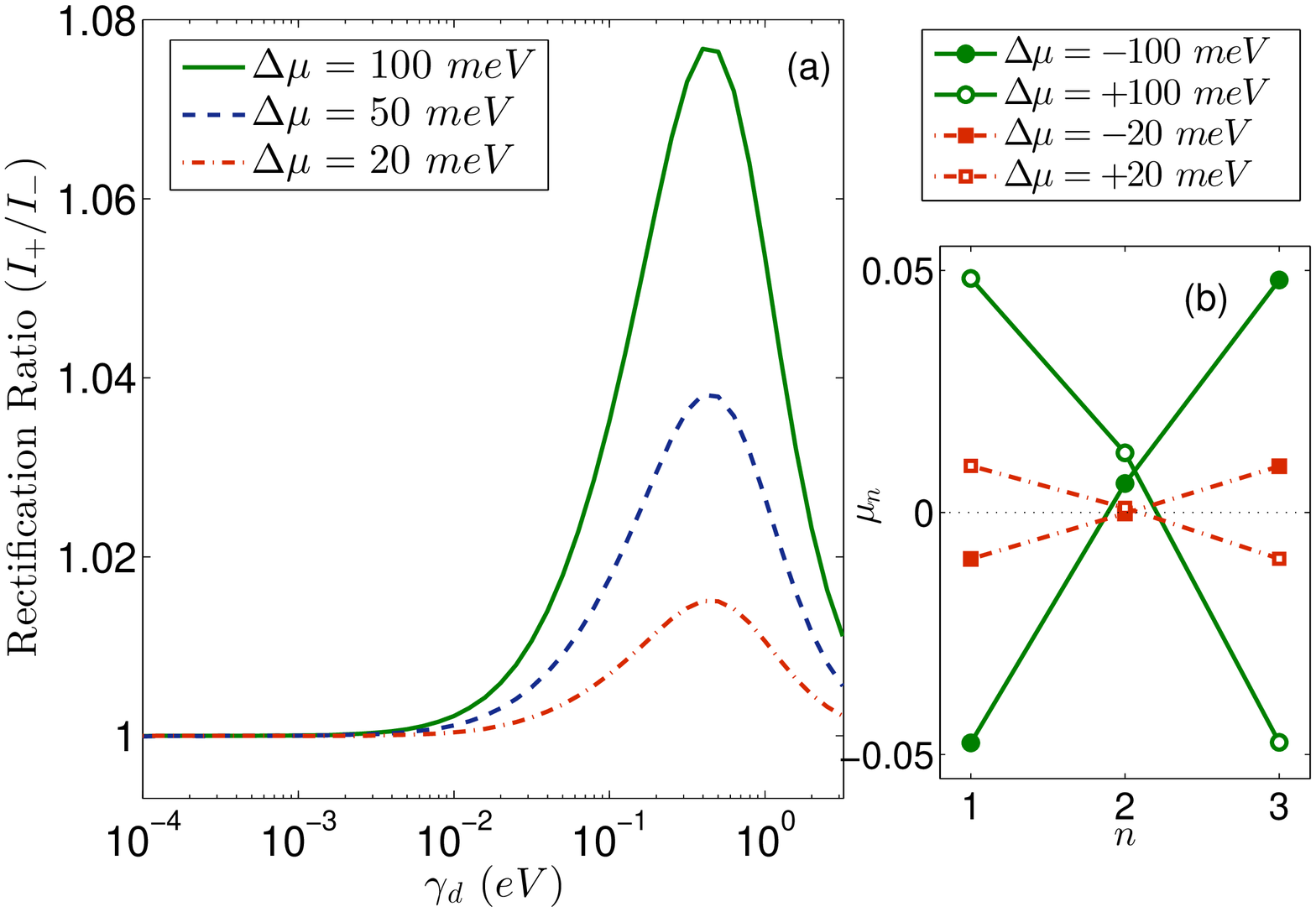}}} 
\caption{
Type-2 EID. 
Top: Schematic representation of the diode.
The electronic structure is spatially symmetric and $\gamma_L=\gamma_R$,  
but coupling strength to the surrounding environment assumes a certain profile, represented by the color gradient of the bath,
$\gamma_{1}> \gamma_2>\gamma_3$, with the integer standing for the site index left to right.
%
(a) Rectification ratio as a function of coupling to the environment.
Parameters are the same as in Fig. \ref{FigD1}, with
$\gamma_1=2\gamma_d$, $\gamma_2=\gamma_d$, $\gamma_3=\gamma_d/2$.
(b) Potential profile within the probes, demonstrating the development of an asymmetry in the junction's response
to applied bias under inelastic effects at peak value $\gamma_d=0.4$ eV.
}
\label{FigD2}
\end{figure}

Fig. \ref{FigD2} illustrates a second type of an EID:
The molecular electronic structure is symmetric, but its coupling to the thermal environment varies monotonically-
asymmetrically along the wire.
Such a modulation could reflect intrinsic molecular-structural asymmetries. Alternatively, it could develop when using
two distinct contacts which induce different local environments on different parts of the molecule \cite{latha15}.
Fig. \ref{FigD2}(a) displays the rectification behavior as a function of $\gamma_d$. Similarly to EID-type 1, we observe an
enhancement of the diode effect with voltage and a
crossover behavior with $\gamma_d$.
The diode effect reflects itself (weakly) in the asymmetric response of the molecule
to inelastic scattering induced by the environment, see Fig. \ref{FigD2}(b). Consistently with  Sec. \ref{SecND},
in linear response situations $\mu_n^++\mu_n^-\sim 0$, missing rectification, while at finite bias
$\mu_n^++\mu_n^->0$, materializing the effect.

We note that the current is higher when $\gamma_d$ is reduced in the same direction of bias drop: $\Delta \mu=\mu_L-\mu_R>0$
and $\gamma_1>\gamma_2>\gamma_3$, than the opposite setup.
This can be rationalized as follows. When $\Delta \mu>0$, high-energy electrons ejected from the $L$ metal
can immediately relax energy on site 1, to physically populate electronic levels. In the reversed direction,
electrons enter the junction to a site weakly interacting with the bath, thus they
minimally dissipate their energy. As a result, electrons cross the junction less effectively, in an off-resonant manner.

\subsection{Absence of rectification with dephasing probes}
\label{SecND}

The EIDs described above operate due to the participation of inelastic electron scattering
in the transport process, implemented with voltage probes.
We prove next that dephasing probes cannot materialize rectification
in these type of models. 
An important working assumption in our derivation, consistent with the discussion in
Secs. \ref{D1}-\ref{D2}, is that molecular levels do not shift under bias.
In what follows, currents and probes' distributions under forward bias ($\mu_L>\mu_R$) are identified by a plus sign;
reversed-bias measures are recorded by a negative sign.

Ignoring physical coefficients, under a forward bias the energy-resolved current leaving the $L$ contact is obtained
from Eq. (\ref{eq:currL}),
\bea
I_L^+(\epsilon)=\mathcal T_{L,R}(\epsilon)[f_L(\epsilon)-f_R(\epsilon)] + \sum_n \mathcal T_{L,n}(\epsilon)[f_L(\epsilon)-f_n^+(\epsilon)],
\nonumber\\
\label{eq:R1}
\eea
with $f^+_n(\epsilon)$ the distribution functions of dephasing probes at each site. These functions are
determined from the following $N$ equations
\bea
I_n^+(\epsilon)&=& \mathcal T_{L,n}(\epsilon)[f_L(\epsilon)-f_n^+(\epsilon)] +
\mathcal T_{R,n}(\epsilon)[f_R(\epsilon)-f_n^+(\epsilon)]
\nonumber\\
&+&
\sum_{n'} \mathcal T_{n',n}(\epsilon)[f_{n'}^+(\epsilon)-f_n^+(\epsilon)]
\nonumber\\
&=&0,
\label{eq:p1}
\eea
see Eq. (\ref{eq:currP}).
We now reverse the bias voltage. The particle current leaving the $L$ contact (now negative) is given by
Eq. (\ref{eq:R1}), only switching $\mu_L$ by $\mu_R$, thus $f_L(\epsilon)$ by $f_R(\epsilon)$,
\bea
I_L^-(\epsilon)=\mathcal T_{L,R}(\epsilon)[f_R(\epsilon)-f_L(\epsilon)] + \sum_n \mathcal T_{L,n}(\epsilon)[f_R(\epsilon)-f_n^-(\epsilon)].
\nonumber\\
\label{eq:R2}
\eea
Note that in general $f_n^-(\epsilon)\neq f_n^+(\epsilon)$,
with  $f_n^-(\epsilon)$ satisfying
the following equations, analogous to Eq. (\ref{eq:p1}),
\bea
I_n^-(\epsilon)&=& \mathcal T_{L,n}(\epsilon)[f_R(\epsilon)-f_n^-(\epsilon)] + \mathcal T_{R,n}(\epsilon)[f_L(\epsilon)- f_n^-(\epsilon)]
\nonumber\\
&+&
\sum_{n'} \mathcal T_{n',n}(\epsilon)[f_{n'}^-(\epsilon)-f_n^-(\epsilon)]
\nonumber\\
&=&0
\label{eq:p2}
\eea
The diode effect corresponds to $\Delta I \equiv I_L^++I_L^- \neq 0$,
a nonzero contribution under bias reversal.
Explicitly, adding up Eqs. (\ref{eq:R1}) and (\ref{eq:R2})
we get
\bea
\Delta I(\epsilon)= \sum_n \mathcal T_{L,n}(\epsilon)[f_L(\epsilon)+f_R(\epsilon)-f_n^+(\epsilon)-f_{n}^-(\epsilon)].
\nonumber\\
\label{eq:Del}
\eea
To prove that rectification is missing, $\Delta I=0$, we follow
a procedure similar to that employed in Ref. \cite{SegalAbs}.
The sum of Eqs. (\ref{eq:p1}) and (\ref{eq:p2}), for every probe $n$, is given by
\bea
&&
[\mathcal T_{L,n}(\epsilon)+\mathcal T_{R,n}(\epsilon)] [f_L(\epsilon)+f_R(\epsilon) - f_n^+(\epsilon)-f_{n}^-(\epsilon)] 
\nonumber\\
&&= -\sum_{n'} \mathcal T_{n',n}(\epsilon)[f_{n'}^+(\epsilon)+ f_{n'}^-(\epsilon) -f_n^+(\epsilon)-f_n^-(\epsilon)].
\label{eq:linear}
\eea
Eq. (\ref{eq:linear})  constitutes a set of $N$ linear equations (which should be solved at every value $\epsilon$).
The $N$ unknowns are the combinations $[f_n^+(\epsilon)+f_{n}^-(\epsilon)]$, and
the inhomogeneous terms are proportional to the nonzero distributions $[f_L(\epsilon)+f_R(\epsilon)]$.
Inspecting Eq. (\ref{eq:linear}), we immediately infer the solution, which is unique,
\bea
f_n^+(\epsilon)+f_{n}^-(\epsilon)
=f_L(\epsilon)+f_R(\epsilon), \,\,\,\, \forall n
\label{eq:solution}
\eea
We substitute this solution into Eq. (\ref{eq:Del}) and receive  $\Delta I=0$.
This concludes our derivation that a diode effect is missing under dephasing probes.

We emphasize that the proof holds for an arbitrary molecular structure, with probes possibly interacting
at different strengths at different sites.
The basic assumptions involved are: (i) The molecular electronic structure stays intact under bias reversal, i.e.  mean field
effects such as level shift under bias are not included.
(ii) Only elastic (yet incoherent) scatterings are allowed via dephasing probes.

How does the proof detailed above fail in the case of voltage probes? Essentially, expressions (\ref{eq:R1})-(\ref{eq:linear})
are valid---after integration over energy.
The analog of equation (\ref{eq:linear}) however becomes non-linear in the case of voltage probes, and rather difficult to solve.
This is because for voltage probes we demand that the functions
$f_n^{\pm}(\epsilon)$ take the form of Fermi functions, thus we need to
solve the equations (\ref{eq:linear}) for the internal variables $\mu_n^{\pm}$.
In this case,  Eq. (\ref{eq:solution}) is not necessarily satisfied, and the junction can rectify current.
At small bias, we perform  Taylor expansions for $f_n^{\pm}$ near equilibrium. In linear response, the probes'
potentials satisfy the symmetry relation $\mu_n^++\mu_n^-\sim \mu_L+\mu_R=0$, and rectification is absent.
The violation of the mirror symmetry $\mu_n^++\mu_n^- \neq 0$ can therefore be associated with the onset of rectification,
as demonstrated in Figs. \ref{FigD1}-\ref{FigD2}.

Classical \cite{Casati-rev} and quantum-mechanical based \cite{SB1,SB2} simulations of charge and heat rectification in nanoscale conductors had demonstrated that two conditions are necessary for the diode effect to set in: The junction should be spatially asymmetric, and many-body effects should participate in conduction. The probe method incorporates phenomenologically many-body electron-phonon effects. The proof above furthermore points out that many-body effects relevant for inducing rectification should involve energy exchange processes.

\section{Conclusions}
\label{Ssum}


We established the applicability of the Landauer-B\"uttiker probe formalism to molecular electronic transport systems under high biases.
This study complements our recent analysis of the LBP method in low-bias transport situations  \cite{Kilgour1}.
The present analysis contributes fundamental understanding of the role played by incoherent elastic and inelastic scattering of electrons in high-bias transport junctions and provides practical-operational directions for designing functionality.

The voltage probe, admitting energy exchange and dissipation processes,
is more relevant to molecular electronic situations than the dephasing probe,
yet the latter is easier to implement.
Our work illustrated that in high bias applications the two probes
may act distinctively, to yield fundamentally different $I$-$V$ characteristics and function.
%
Specifically, one of our main observations was that under large bias, when molecular electronic levels are dispersed from each other
(yet placed within the bias window), elastic decoherence effects
degrade the current with bias, while energy exchange processes facilitate conduction.

Concerning applications, designs for molecular diodes are based at large on the assumption of pure coherent motion,
with dephasing-incoherent effects degrading performance.
Here, in contrast, we identified two types of diodes which operate only due to the
interaction of electrons with their thermal environment.
We further provided a rigorous proof that dephasing probes cannot support charge-current rectification without additional effects,
e.g., mean-field level-shift under bias.

In summary, the phenomenological probe technique for simulating incoherent effects in electronic conduction is gaining applications in molecular
transport junctions. Our work illustrates that: (i) At high bias, dephasing and voltage probes may predict significantly different $I$-$V$ curves.
(ii) Experimental results on linear chains, showing the enhancement of current with voltage at high voltage $\Delta \mu\gg \epsilon_B$, are more consistently described with the (more difficult to implement) voltage probe, evincing on the important role of inelastic effects in transport.
(iii) Incoherent-inelastic effects combined with structural asymmetry may yield small rectification.
In future works we will study the role of inelastic effects on other molecular electronic phenomena, using the probe method:
cooperative effects \cite{Arik,Reuter-nano11,Reuter-nano12} and molecular thermoelectric performance. 

%

%

%

\begin{acknowledgments}
The work was supported by the Natural Sciences and Engineering Research Council of Canada and the Canada Research Chair Program.
The work of Michael Kilgour was partially funded by an Ontario Graduate Scholarship.
\end{acknowledgments}



\begin{thebibliography}{32}
\bibitem{pathway}
Numerical simulations \cite{Nozaki08,Nozaki12,Pastawski-poly,Chen-Ratner,WaldeckF,Kilgour1,Anantram_linear}
and analytical studies (mostly for heat conduction)
\cite{Dhar,Bonetto,DRoy,Roy}
had demonstrated the emergence of incoherent conduction due to the coupling of local sites to either dephasing or voltage probes.
This could be reasoned in terms of the energy eigenstates of the molecule, $\hat H_M|M\rangle =E_M|M\rangle$.
The probes scatter electrons in an incoherent manner, elastically or inelastically, between different electronic eigenstates $|M\rangle$,
introducing an incoherent transport mechanism on top of the coherent (direct left-to-right transmission) contribution.
The hybridization of eigenstates to the contacts (at least to one of them) decays exponentially with size in off-resonant situations, leading to small tunneling conduction. The probes however open up a more effective charge transfer pathway
between the state $|M\rangle$,
which is strongly coupled to the left lead, and state $|M'\rangle$, which strongly couples to the right terminal.
In long chains of length $N$, the most effective pathway involves multiple scattering between eigenstates, with
the conductance scaling approximately with $N^{-1}$ \cite{Kilgour1}.
\end{thebibliography}

\end{document}